\documentclass{article}
\begin{document}
\baselineskip18pt
\title{Shannon versus Kullback-Leibler Entropies in Nonequilibrium Random Motion}
\author{Piotr Garbaczewski\thanks{Corresponding author:
p.garbaczewski@if.uz.zgora.pl}\\
 Institute of Physics,  University
of Zielona G\'{o}ra, PL 65-516 Zielona  G\'{o}ra, Poland}
\maketitle
\begin{abstract}
We  analyze dynamical properties of   the Shannon information
entropy of a continuous probability distribution, which is driven by  a
standard diffusion process. This entropy  choice is confronted  with another  option,
 employing the conditional Kullback-Leibler entropy.  Both entropies discriminate
  among   various  probability   distributions, either statically or in the
 time domain.  An asymptotic approach towards equilibrium  is typically  monotonic in
 terms of the Kullback entropy. The Shannon entropy time rate  needs not to be positive
 and is a  sensitive   indicator of  the power transfer processes (removal/supply)  due to
  an active environment. In  the case of  Smoluchowski diffusions,
    the Kullback  entropy  time  rate  coincides with the   Shannon  entropy "production" rate.
\end{abstract}

\vskip0.3cm

 We  consider diffusion
processes in $R$  with a  constant or time-dependent diffusion
coefficient $D$, characterized  (non-uniquely)  by space-time
inhomogeneous probability densities $\rho = \rho (x,t)$  and local
velocity fields  (drifts or currents). We specify the  diffusive
dynamics  to refer to of a mass $m$ particle in the external field
of force, here taken to  be conservative: $F=F(x) =-{\nabla } V$.
 The associated  Smoluchowski diffusion process
with a forward drift ${b}(x) = \frac{F}{m\beta }$ is analyzed in
terms of the   Fokker-Planck equation for the probability
density $\rho(x,t)$, \cite{risken,hasegawa,rubi}:
\begin{equation}
\partial _t\rho =
D\triangle \rho -  {\nabla } (b\cdot  \rho )\,  \label{Fokker}
\end{equation}
with the initial data $\rho _0(x) = \rho (x,0)$.
We assume   a priori the Einstein fluctuation-dissipation relationship
 $D=\frac{k_BT}{m\beta }$, where  $\beta $ is
interpreted as  the friction (damping) parameter,  $T$ is the
temperature of the bath, $k_B$ being the Boltzmann constant .

 With a solution  $\rho (x,t)$ of  the Fokker-Planck equation, we
  associate the  Shannon information entropy of a continuous probability distribution,
   also named  differential  entropy, \cite{shannon,cover,sobczyk}:
\begin{equation}
{\cal{S}}(t)=  - \int \rho (x,t)\, \ln \rho (x,t) \,  dx
\end{equation}
 which   is typically
\it not \rm a conserved quantity. The rate of change in time of
${\cal{S}}(t)$   readily follows.

Anticipating further discussion, let us stress that  even in case of
 plainly irreversible diffusive  dynamics, it is by no means obvious  that
  the differential entropy should grow,  decay (diminish) or show up a mixed behavior.
 Normally one tries, in terms of a suitable entropy  definition,  to justify the  popular
 albeit tacit
assumption  that "generically"  the time rate of  "entropy" (whatever it is, be that
$\dot{\cal{S}}>0$) is non-negative.

We take for granted the validity of standard assumptions which allow to interchange derivatives
with indefinite integrals. Boundary restrictions  upon  $\rho $, $v \rho $ and $b \rho $ to
vanish at spatial infinities (or  at finite spatial volume
boundaries) yield the  rate  equation:
\begin{equation}
{\frac{d{\cal{S}}}{dt}}  = \int [\rho \, (\nabla \cdot b)
 + D \cdot  {\frac{(\nabla \, \rho )^2}\rho }]\, dx  \label{force}
\end{equation}
which in case of  $b =0$  refers to the one-dimensional free Brownian motion
(Wiener process), when    $\dot{\cal{S}}   >0$.
The trivial example is provided by the familiar  heat kernel
$\rho (x,t) = (4\pi Dt)^{-1/2} \exp ( -x^2/4Dt) $  whose  differential entropy  reads
${\cal{S}}(t) =  (1/2) \ln (4\pi e Dt)$, implying  $\dot{\cal{S}} =1/2t >0$.

 We can  rewrite  Eq. (\ref{force})  in  a number of equivalent ways.  To this end let us
introduce the notion of   the  current velocity of Brownian
particles in external force field (with a short-hand notation $v
\doteq v(x,t)$): $v \doteq   b - u
 = \frac{%
 F}{\ m\beta }- D\frac{\nabla
\rho }{\rho }$
which  allows us  to   transform the
 Fokker-Planck equation into  the continuity equation $\partial _t\rho = -
  \nabla (v\cdot \rho )$  and back.
Since  $\langle u^2 \rangle = - D \langle \nabla \cdot u \rangle
$, we have: $ D \dot{\cal{S}} \doteq  D \left<
{\nabla }\cdot
 {b} \right>  +  \left< {u}^2\right> =  D\langle \nabla \cdot
 v\rangle   = -  \langle {v}\cdot
  {u} \rangle $.

By reasons to become clear in below, we shall  pay  particular attention to the
  entropy balance equation  in the form:
\begin{equation}
D \dot{\cal{S}}  =  \langle {v}^2\rangle
    -  \langle b\cdot {v}  \label{balance}
 \rangle
 \end{equation}
where $\langle  \cdot  \rangle $ denotes the mean value with
respect to $\rho $.

This  balance  equation  is extremely   persuasive when recast in the thermodynamically
motivated form:
\begin{equation}
\frac{d{\cal{S}}}{dt} = \left( {\frac{d{\cal{S}}}{dt}}\right)_{in}
- {\frac{d{\cal{Q}}}{dt}}\, . \label{balance1}
\end{equation}

The  non-negative term  $(\dot{\cal{S}})_{in} = (1/D) \langle v^2\rangle $ in
Eq.~(\ref{balance}) may  be  interpreted as   the measure of time  rate at which  the
Shannon information  entropy of the process  shows a tendency   to grow   (is generated due to an
implicit thermal agitation), at the  expense  of the environment.
 This  excess  "entropy  production"   \it by \rm  the environment    may  possibly  be
     counterbalanced by the entropy/heat removal due to dissipation,   provided we have  $\dot{{\cal{Q}}} >0$.
The   characteristic
"power release" expression:
\begin{equation}
{\frac{d{\cal{Q}}}{dt}} \doteq  {\frac{1}D} \int {\frac{1}{m\beta }}
 {F} \cdot  {j}\,  dx = {\frac{1}D} \left\langle b\cdot v
 \right\rangle \,  , \label{heat1}
\end{equation}
in the formal thermodynamical lore is  expected to
 refer to the time rate at which the mechanical work per unit of mass is
 dissipated  (returned back  to the reservoir) in the form of heat.
 Notice that  $k_BT \dot{{\cal{Q}}}= \int F\cdot j\,  dx$, where $b= F/(m\beta )$,
   $j = v \rho $ and   $T$ stands for   the temperature of the bath.

However, in the diffusion context,  $\dot{\cal{Q}}$  has an  unspecified sign. It  may be positive,
with an interpretation of the power  absorption by  the environment. $\dot{\cal{Q}}$  may take
negative values, when the power is drained out from the  environment.
In below we shall discuss those issues for the exemplary
Ornstein-Uhlenbeck process.

Of particular interest  is the case of a constant information
entropy $\dot{\cal{S}} =0$ which amounts to  the existence of
steady states, \cite{qian}. In the simplest case, when the
 diffusion current  vanishes, we  encounter the primitive  realization of the state of  equilibrium
  with an invariant density $\rho $. Then,    $b=u = D \nabla  \ln \rho  $
 and we  readily  arrive at the classic  equilibrium identity  for the Smoluchowski process:
\begin{equation}
-(1/k_BT)\nabla V = \nabla \ln\, \rho  \, \label{eq}
\end{equation}
which determines the functional form of the invariant density  in
case of  a given conservative force field, \cite{risken,lasota}.
There is an ample discussion in  Ref.~\cite{qian}
 of how these properties match with  time reversal of the stationary diffusion process and the
 vanishing  of  the entropy  production  rate  $(\dot{\cal{S}})_{in}$.

Coming back to the general discussion, let us define the so-called (fictitious)
thermodynamic force $F_{th}\doteq  v/D$  associated with the
Smoluchowski diffusion  and introduce its  corresponding
time-dependent
 potential function $\Psi (x,t)$:
 \begin{equation}
 k_BT\, F_{th}  = F - k_BT\, \nabla  \ln \rho
 \doteq - {\nabla } \Psi  \, .
  \end{equation}

Notice that   $v=  - (1/m\beta ) \nabla \Psi $. In the absence of
external forces (free Brownian motion), we obviously get $F_{th} =
- \nabla \ln \rho = - (1/D) u $.

The  mean value of the potential
\begin{equation}
\Psi = V + k_BT \ln \rho  \label{Helmholtz}
\end{equation}
associates  with the diffusion process  an obvious  analogue of the
 thermodynamical  Helmholtz free energy:
\begin{equation}
\left< \Psi \right> = \left< V\right> - T\, {\cal{S}}_G
\end{equation}
where the dimensional (Gibbs-type)  version ${\cal{S}}_G  \doteq k_B {\cal{S}}$
of the  information entropy has been introduced. The
expectation value  of the mechanical force potential $ \left<
V\right>$ plays here the role of (mean) internal energy.

By assuming that $\rho V {v}$  vanishes at integration volume
boundaries (or infinity), we easily get the time rate of Helmholtz
free energy at a constant temperature $T$:
\begin{equation}
{\frac{d}{dt}} \left< \Psi  \right> = -  k_BT \dot{{\cal{Q}}}  - T
\dot{\cal{S}}_G \, . \label{temp}
\end{equation}

By employing Eq.~(\ref{balance1}),  we readily  arrive at
\begin{equation}
 {\frac{d}{dt}} \left< \Psi  \right>  = -  (k_BT ) \left( {\frac{d{\cal{S}}}{dt}}\right)_{in}  = - (m\beta )
 \left< {v}^2\right>  \label{growth}
\end{equation}
which   either identically vanishes (equilibrium)  or remains
negative.

 Thus, Helmholtz free energy either remains constant in time
or decreases as a function of time at  the rate set by the
information entropy "production"   $\dot{\cal{S}}_{in}$. One may
expect that actualy $\langle \Psi \rangle (t)$ drops down to a
finite minimum as $t\rightarrow \infty $.

 However, this feature is a little bit  deceiving. One should be aware that
 a finite minimum  as well may  not exist, which is the case e.g.  for the
  free Brownian motion.  Also, non-unique  minima need to be excluded as
  well.

Till now,   we  have deliberately
avoided any use  of the relative  Kullback-Leibler entropy,
\cite{cover,sobczyk,lasota}, which  is often
invoked  for a comparison purpose  to tell "how far from each other"  two probability
densities  are.
However,  a  reliability  of the Kullback entropy   may be
questioned:  this entropy  fails to quantify properly a "comparison" of once given
density function to itself, if   considered  at different instants of
its   non-stationary time evolution.

To  analyze  defective  features of the Kullback entropy in a
fully controllable  way, let us consider
a one parameter family  of  Gaussian densities   $\rho _{\alpha  } = \rho (x-\alpha  )$,
   with the mean  $\alpha  \in R$ and
 the  standard deviation fixed at $\sigma $.  They    share    the  very same value  $ {\cal{S}}_{\sigma }
 =   {\frac{1}{2}} \ln \,(2\pi e \sigma ^2)$ of the  Shannon information  (differential)  entropy,
 independent of $\alpha $.

If we admit $\sigma $ to be an additional  free parameter, a
two-parameter family of Gaussian densities
  $\rho _{\alpha  } \rightarrow \rho _{\alpha , \sigma }(x)$ appears.  Such
  densities,  corresponding to   different  values of  $\sigma $ and $\sigma '$ do
admit an "absolute comparison" in terms of the Shannon entropy:
\begin{equation}
{\cal{S}}_{\sigma '} - {\cal{S}}_{\sigma } =  \ln \, \left(
\frac{\sigma '}{{\sigma }}\right) \,  \label{comparison}
\end{equation}
and the outcome is insensitive to translation parameters $\alpha $  and $\alpha '$.

By denoting    $\sigma \doteq \sigma (t)$ and $\sigma
' \doteq \sigma (t')$ we  can  make any
density  amenable to the "absolute comparison"  formula
 at different time instants $t'>t >0$. For the heat kernel we have $\sigma (t) = \sqrt{2Dt}$
 and therefore $(\sigma '/ \sigma ) = \sqrt{t'/t}$.

There are many inequivalent ways to evaluate the "divergence" or "convergence"  between probability
distributions. The  relative (Kullback) entropy is typically used
to quantify such   divergence  relative to  the  prescribed
reference density, \cite{lasota}.

We define the  Kullback  entropy ${\cal{K}}(\theta ,\theta')$ for
a one-parameter family of probability densities $\rho _{\theta}$,
so that the   "distance" between  any two densities in this family
can be directly  evaluated.  Let  $\rho _{\theta '}$ denote
the reference probability density. We have, \cite{cover,sobczyk}:
\begin{equation}
{\cal{K}}(\theta ,\theta ') \doteq  {\cal{K}}(\rho _{\theta }|\rho
_{\theta '}) =    \int \rho _{\theta }(x)\, \ln {\frac{\rho
_{\theta }(x)}{\rho _{\theta '}(x)}}\, dx \, .
\end{equation}
which, in view of the concavity of the function $f(w) = - w \ln
w$,  is  positive.

 Let us indicate that the negative of
${\cal{K}}$, ${\cal{H}}_c \doteq - {\cal{K}}$,    named the
conditional entropy, is predominantly used in the
literature \cite{sobczyk,tyran,lasota}  because of  its  affinity
(regarded as a  formal  generalization)  to the differential
entropy. Then e.g. one investigates an approach of $-{\cal{K}}$
towards its maximum (usually achieved at the value zero) when  a
running density is bound to have a  unique stationary asymptotic,
\cite{tyran}.

If we take  $\theta ' \doteq \theta + \Delta \theta$ with $ \Delta
\theta  \ll 1$,  the following approximate formula holds true
under a number of  standard assumptions:
\begin{equation}
{\cal{K}}(\theta ,\theta + \Delta  \theta ) \simeq {\frac{1}{2}}
{\cal{F}}_{\theta }\, \cdot  (\Delta \theta )^2  \label{approx}
\end{equation}
where ${\cal{F}}_{\theta }$  denotes so-called  Fisher information
measure.  With this
proviso, we can  evaluate the Kullback distance within
 a two-parameter $(\alpha , \sigma )$ family of Gaussian
densities, by taking $\theta \rightarrow \alpha $.

Passing to $\alpha  '= \alpha  +\Delta \alpha  $ at  a fixed value
of  $ \sigma $  we arrive at:
\begin{equation}
{\cal{K}}(\alpha ,\alpha  + \Delta  \alpha  ) \simeq
{\frac{(\Delta \alpha )^2} {2\sigma ^2}}\, .
\end{equation}
For the record, we  note   that  the  respective Shannon entropies
do  coincide: ${\cal{S}}_{\alpha } = {\cal{S}} _{\alpha + \Delta
\alpha }$.

Analogously, we can  proceed with respect to the label $\sigma $
at  $\alpha $ fixed:
\begin{equation}
{\cal{K}}(\sigma , \sigma + \Delta \sigma ) \simeq {\frac{(\Delta
\sigma )^2}{ \sigma ^2}}
\end{equation}
when, irrespective of  $\alpha $:
\begin{equation}
{\cal{S}}_{\sigma + \Delta \sigma } - {\cal{S}}_{\sigma } \simeq
{\frac{\Delta \sigma } {\sigma }}\, .
\end{equation}

By choosing  $\theta \rightarrow \sigma ^2$ at  $\alpha $ fixed,
we get (now the variance $\sigma ^2$ is modified by its increment
$\Delta (\sigma ^2)$):
\begin{equation}
{\cal{K}}(\sigma ^2 , \sigma ^2 + \Delta (\sigma ^2)) \simeq
{\frac{[\Delta (\sigma ^2)]^2}{4 \sigma ^4}}
\end{equation}
while
\begin{equation}
{\cal{S}}_{\sigma ^2 + \Delta (\sigma ^2) } - {\cal{S}}_{\sigma
^2} \simeq {\frac{\Delta (\sigma  ^2)}{2\sigma ^2 }}
\end{equation}
which, upon  identifications  $\sigma ^2 = 2Dt$ and $\Delta
(\sigma ^2)= 2D\Delta t$, sets an obvious connection with the
 differential $(\Delta {\cal{S}})(t)$ and thence   with  the
  time derivative $\dot{\cal{S}} = 1/2t$ of the heat kernel
differential entropy.

Our  previous observations are a special case of more general
reasoning. Namely, if we consider  a two-parameter $\theta \doteq
(\theta _1,\theta _2)$  family of densities, then instead of
Eq.~(\ref{approx}) we would have  arrived at
\begin{equation}
{\cal{K}}(\theta ,\theta + \Delta  \theta ) \simeq {\frac{1}{2}}
 \sum_{i,j} {\cal{F}}_{ij}\, \cdot  \Delta \theta _i  \Delta \theta _j
\end{equation}
where $i,j, = 1,2$ and the Fisher information matrix
${\cal{F}}_{ij}$ is given as follows
\begin{equation}
{\cal{F}}_{ij} = \int \rho _{\theta } {\frac{\partial \ln \rho
_{\theta } }{\partial \theta _i}}  \cdot {\frac{\partial \ln \rho
_{\theta } }{\partial \theta _j}}\, dx \, .
\end{equation}

In case of Gaussian densities, labelled  by  independent
parameters $\theta _1 = \alpha $ and $\theta _2 = \sigma $
(alternatively $\theta _2 = \sigma ^2$), the Fisher matrix is
diagonal and defined in terms of previous entries
${\cal{F}}_{\alpha }$ and ${\cal{F}}_{\sigma }$ (or
${\cal{F}}_{\sigma ^2}$).

It is useful  to  note  (c.f. also \cite{tyran})  that in
self-explanatory notation, for  two $\theta $ and $\theta '$
Gaussian densities there holds:
\begin{equation}
{\cal{K}}(\theta ,\theta ')=  \ln {\frac{\sigma '}{\sigma }} +
{\frac{1}{2}}({\frac{\sigma ^2}{{\sigma '}^2}} - 1) +
{\frac{1}{2{\sigma '}^2}} (\alpha - \alpha ')^2  \label{kulgau}
\end{equation}
The first entry in Eq.~(\ref{kulgau}) coincides with the
"absolute comparison formula" for Shannon entropies,
Eq.~(\ref{comparison}). However for $|\theta ' - \theta |\ll 1$,  the second term dominates
the first one.

Indeed, let us set $\alpha ' = \alpha $ and consider $\sigma ^2 =
2Dt$, $\Delta (\sigma ^2)= 2D\Delta t$. Then ${\cal{S}}(\sigma ')
- {\cal{S}}(\sigma ) \simeq \Delta t/2t$, while ${\cal{K}}(\theta
,\theta ') \simeq (\Delta t)^2/ 4t^2$. Although, for finite
increments $\Delta t$ we have
\begin{equation}
 {\cal{S}}(\sigma ') -
{\cal{S}}(\sigma )\simeq \sqrt{ {\cal{K}}(\theta ,\theta ')}\simeq
{\frac{\Delta t}{2t}} \, ,
\end{equation}
the time derivative  can be defined exclusively   for the differential entropy, $\dot{\cal{S}}$,
 and is meaningless  in terms of  the Kullback "distance".

 Let us mention that no such obstacles arise  in the standard cautious  use of the
 relative  Kullback entropy ${\cal{H}}_c$.  Indeed, normally one
 of the involved densities stands for the  stationary  reference one
 $\rho _{\theta '}(x) \doteq \rho _*(x)$,  while  another evolves in time
 $\rho _{\theta }(x) \doteq \rho (x,t)$, $t\in R^+$, thence we need:
 \begin{equation}
 {\cal{H}}_c(t) \doteq - {\cal{K}}(\rho _t|\rho _*) \, .
\end{equation}

 In the presence of external forces,   the property Eq.~(\ref{growth})  may consistently  quantify
 an asymptotic approach towards a minimum corresponding to an invariant (presumed
to be unique) probability  density of the process. Indeed, by
invoking Eq.~(\ref{eq}) we realize that
\begin{equation}
\rho _*(x)= {\frac{1}{Z}} \exp \left(-{\frac{V(x)}{k_BT}}\right)
\end{equation}
 where $Z=\int \exp(-V(x)/k_BT)\, dx $, sets
the   minimum   of  $\langle \Psi \rangle (t)$ at $\langle \Psi
\rangle _* = \Psi _*= -k_BT \ln  Z$.

Let us take the above $\rho _*(x)$  as a reference density  with
respect to which the divergence of $\rho (x,t)$ is evaluated in
the course of the  pertinent Smoluchowski process. This divergence
is well quantified  by the conditional  Kullback entropy
${\cal{H}}_c(t)$,   where:
\begin{equation}
{\cal{H}}_c(t) =  -  \int     \rho \, \ln \left({\frac{\rho }{\rho
_*}}\right )\, dx  =  {\cal{S}}(t) - \ln Z - {\frac{\langle
V\rangle }{k_BT}}\, .
\end{equation}

Consequently, in view of Eqs.~(\ref{temp})  and (\ref{balance1}),
we get
\begin{equation}
\dot{\cal{H}}_c = \dot{\cal{S}} + \dot{\cal{Q}} =
(\dot{\cal{S}})_{in} \geq 0
\end{equation}
so   that ${\frac{d}{dt}} \langle \Psi \rangle =  - (k_BT)\,
\dot{\cal{H}}_c$. An approach of $\langle \Psi \rangle (t)$
towards the minimum  proceeds in the very same rate as this of
${\cal{H}}_c(t)$   towards its maximum.

In contrast to $\dot{\cal{H}}_c$ which is non-negative, we have no
growth guarantee for the differential entropy, since the sign of  $\dot{\cal{S}}$
 is unspecified. Nonetheless, the balance between the
time  rate of entropy  production/removal   and the power release
into or out of the  environment, is  definitely correct and   informative.

 We have  $\dot{\cal{S}} \geq - \dot{\cal{Q}}$  and  surely
  $\dot{\cal{Q}}<0$ $\rightarrow $ $\dot{\cal{S}} >0$.  If $\dot{\cal{Q}}>0$,
$\dot{\cal{S}}$ may take negative values  down to the lower bound $- \dot{\cal{Q}}$.

It is quite illuminating to exemplify previous considerations  by
a detailed presentation of the standard one-dimensional
Ornstein-Uhlenbeck process. We denote $b(x)= - \gamma x  $ with
$\gamma >0$ and choose  an initial density  in the Gaussian form, with the
mean value $\alpha _0$ and variance $\sigma ^2_0$. The
Fokker-Planck evolution  Eq.~(\ref{Fokker})  preserves the
Gaussian form  of $\rho (x,t)$ while modifying
 the mean value and variance according to  $\alpha (t) = \alpha _0 \exp(-\gamma t)$, and
\begin{equation}
 \sigma ^2(t) = \sigma ^2_0 \exp (- 2 \gamma t) + {\frac{D}{\gamma }}[1-\exp (-2\gamma t)] \, .
\end{equation}

Accordingly, since   a unique invariant density has the form $\rho
_* =   \sqrt{\gamma /2\pi D}
 \exp (-\gamma x^2/2D)$ we obtain, \cite{tyran}:
\begin{equation}
{\cal{H}}_c(t) = \exp (-2 \gamma t) {\cal{H}}_c(\rho _0,\rho _*)=
-{\frac{\gamma \alpha _0^2}{2D}}\, \exp (-2\gamma t)
\end{equation}
while in view of our previous considerations, we have
${\cal{S}}(t) =(1/2) \ln [2\pi e \sigma ^2(t)] $ and ${\cal{F}} =
1/\sigma ^2(t)$.  Therefore
\begin{equation}
\dot{\cal{S}}= {\frac{2\gamma (D - \gamma \sigma _0^2) \exp(-2
\gamma t)} {D -(D - \gamma \sigma _0^2)\exp(-2\gamma t)}}\, .
\end{equation}

We observe that  $\sigma ^2_0 > D/\gamma  \rightarrow \dot{\cal{S}}
<0$, while $\sigma ^2_0 < D/\gamma \rightarrow \dot{\cal{S}} > 0$.
In both cases the behavior of the differential   entropy is
monotonic, although its   ultimate  growth or decay do critically rely on  the
choice of $\sigma ^2_0$. Irrespective of $\sigma ^2_0$ the
asymptotic value of   ${\cal{S}}(t)$ as  $t\rightarrow \infty $
reads $(1/2) \ln [2\pi e (D/\gamma )$.

The differential entropy evolution is anti-correlated with  the Fisher  measure
 of the  probability  localization, since
\begin{equation}
\dot{\cal{F}}= -\,  {\frac{\gamma \dot{\cal{S}}} {[D -(D - \gamma
\sigma _0^2)\exp(-2\gamma t)]^2}}\, .
\end{equation}
For all $\sigma _0^2$  the asymptotic value of ${\cal{F}}$  reads
$\gamma /D$.

We have here a direct control of the behavior of the "power
release" expression $\dot{\cal{Q}} = \dot{\cal{H}}_c -
\dot{\cal{S}}$.  Since
\begin{equation}
\dot{\cal{H}}_c= (\gamma ^2\alpha _0^2/D) \exp(-2\gamma t) >0\, ,
\end{equation}
in case of $\dot{\cal{S}} <0$ we encounter
 a continual power removal  $\dot{\cal{Q}}> 0$  into  the thermal environment.

 In case of $\dot{\cal{S}} >0$ the situation is more complicated. For example, if $\alpha _0 =0$, we
  can easily check that  $\dot{\cal{Q}}  < 0$, i.e. we have the power  supply  from   the
  environment  for all $t\in R^+$.
  More generally, the sign of  $\dot{\cal{Q}}$ is
  negative for   $\alpha _0^2< 2(D-\gamma \sigma ^2_0)/\gamma$. If the latter inequality  is reversed, the sign of
  $\dot{\cal{Q}}$ is not uniquely specified  and suffers  a change at a suitable time  instant $t_{change}
  (\alpha _0^2,\sigma _0^2)$.

Standard notions of thermodynamical  entropy are basically not  considered  in the time domain. However
any conceivable idea of "approaching" the state of equilibrium or passing from one such state to another (steady) state,
surely involves the time dependence and the related non-equilibrium dynamical process. This refers to
 attempts to give a precise meaning to Boltzmann's H-theorem under  non-equilibrium conditions and search for
 an origin of  increasing entropy in terms of model systems. Consult e.g. at this point,  both  standard motivations
and apparent  problems encountered in connection with the
$H$-theorem  and its diffusion process analogues,  \cite{hasegawa,qian,lasota}.

 Our   analysis of simple diffusion-type  models indicates that the  very  notion
 of entropy is  non-universal and  purpose-dependent. In particular, although the conditional
 Kullback  entropy is often considered as as  the  only valid "entropy growth"  justification,
 a deeper  insight into the underlying physical phenomena  (power transfer processes)  is available only
through  the differential (Shannon)  entropy notion.

{\bf  Acknowledgement:} The paper has been supported by the Polish Ministry of Scientific Research
an Information Technology (solicited) grant No PBZ-MIN-008/P03/2003.

\end{document}